\documentclass[showpacs,preprintnumbers,amsmath,amssymb]{revtex4}

\usepackage{graphicx}
\usepackage{dcolumn}
\usepackage{bm}
\usepackage{float}

\newcommand{\ket}[1]{\vert #1 \rangle}

\begin{document}

\title{Electromagnetically induced transparency and four-wave mixing in a cold atomic ensemble with large optical depth}

\author{J.~Geng$^{1,2}$, G.~T.~Campbell$^2$, J.~Bernu$^2$, D.~Higginbottom$^{2}$, B.~M.~Sparkes$^2$, S.~M.~Assad$^2$, W.~P.~Zhang$^1$, N.~P.~Robins$^3$}
\author{P.~K.~Lam$^{2,4}$}
\email{Ping.Lam@anu.edu.au}
\author{B.~C.~Buchler$^2$}%
\email{Ben.Buchler@anu.edu.au}
\address{$^1$Department of Physics, State Key Laboratory of Precision Spectroscopy,
East China Normal University, Shanghai 200062, P.~R.~China}
\address{$^2$Centre for Quantum Computation and Communication Technology, Department of Quantum Science, The Australian National University, Canberra, ACT 0200, Australia}
\address{$^3$Quantum Sensors Lab, Department of Quantum Science, The Australian National University, Canberra, ACT 0200, Australia}
\address{$^4$College of Precision Instrument and Opto-electronics Engineering, Key Laboratory of Optoelectronics Information Technology of Ministry of Education, Tianjin University, Tianjin, 300072, P.~R.~China}

\date{\today}

\begin{abstract} 

We report on the delay of optical pulses using electromagnetically induced transparency in an ensemble of cold atoms with an optical depth exceeding 500. To identify the regimes in which four-wave mixing impacts on EIT behaviour, we conduct the experiment in both Rb$^{85}$ and Rb$^{87}$. Comparison with theory shows excellent agreement in both isotopes. In Rb$^{87}$, negligible four-wave mixing was observed and we obtained one pulse-width of delay with 50\% efficiency. In Rb$^{85}$ four-wave-mixing contributes to the output. In this regime we achieve a delay-bandwidth product of 3.7 at 50\% efficiency, allowing temporally multimode delay, which we demonstrate by compressing two pulses into the memory medium.

\end{abstract}

\maketitle

\section{Introduction}
Electromagnetically induced transparency  (EIT)  \cite{Marangos1998_EITReview,Fleischhauer:RevEIT:2005} is a coherent atom-optical effect that arises due to quantum interference of optical transitions. Since its first experimental observation in a strontium vapour in 1991 \cite{Boller:1991if}, it has been investigated in numerous atomic systems in a wide variety of settings. It is of fundamental interest for its ability to slow light by up to seven orders of magnitude \cite{Budker:1999hd, Hau:1999cz}. In turn, the increased interaction times afforded by slow light allow enhanced non-linear optical interactions \cite{Harris:1990ey,Jain:1996vk,Harris:1999vj}.  EIT can also be used to stop and store light in an atomic spinwave leading to its application as a quantum memory \cite{PhysRevA.65.022314,Alex:PRL:2008:100,Honda:2008p4680,Kimble_nature_sph-QM1,Tittel2009_MemReview}.  In order to preserve a quantum state, a quantum memory requires storage that is both efficient and noiseless. Ideally, EIT has the potential to be used as a high fidelity quantum memory \cite{Hetet:2008dm}. Considerable work has been done to improve the efficiency of EIT memories, primarily by increasing the optical depth of the slow-light medium. Recent results have demonstrated up to 69\% efficiency for storage and forward recall of light \cite{Chen2013}. In principle, the achievable storage efficiency using EIT can approach 100\% \cite{Gorshkov2007} as the optical depth is made sufficiently large. In this regime the delay of an optical pulse travelling through the EIT medium can be larger than the duration of the pulse \cite{PhysRevA.71.023801} such that the pulse is contained entirely within the medium.

At large optical depths, however, nonlinear processes such as four-wave mixing (4WM) may become significant. In particular, these processes may introduce gain during slow-light propagation that may contribute noise to the output \cite{PhysRevA.78.023801} \cite{PhysRevA.88.013823}, thus diminishing the suitability of EIT as a quantum memory. Experimental work in warm atomic vapours has shown that EIT may have significant 4WM \cite{Phillips2009a,Phillips2011}. In principle, the unbroadened atomic states available in cold atomic ensembles allows high storage efficiency with little added noise due to having a lower 4WM strength than their warm counterparts. Moreover, it has also ben reported that the long atomic coherence in cold system also delivers high storage efficiencies \cite{Chen2013}. 

In this paper, we report on EIT in an ensemble of cold $^{85}Rb$ atoms with a very high optical depth. We investigate the role of 4WM in the experiment and demonstrate that 4WM can have a non-negligible contribution to the intensity of the signal field after EIT delay. By repeating the experiment in cold $^{87}Rb$ atoms, which have larger detuning between the two ground states, we show that the 4WM can be diminished to the point of having negligible contribution to the signal field at comperable optical depths. Good agreement between our results and theory indicates that the theoretical model used in a recent paper by Lauk {\it et al.} \cite{PhysRevA.88.013823} is a realistic description of cold atom EIT.

We also show that with sufficiently high optical depth, delays of more than two pulse-widths are possible. This is a step towards a temporally multimode quantum memory for increasing the success rate for quantum information protocols \cite{Simon:2007}.  Previous experiments have demonstrated large fractional delays using the strong dispersion of dense optical vapours \cite{Camacho2007}. In this paper, we present EIT results with sufficient delay to store multimode pulses.

In the next section, we begin by reviewing the semi-classical theory of EIT-4WM presented in Ref.~\cite{PhysRevA.88.013823}. We then use this theory to model our system.  In section \ref{sec-setup}, we present an overview of the cold atom trap and the timing sequence scheme used to prepare the atomic ensemble. We discuss the experimental results in section \ref{sec-results} before concluding in section \ref{sec-concl}. 

\section{Theory}

\begin{figure}
\centering \includegraphics[width= 0.4\columnwidth]{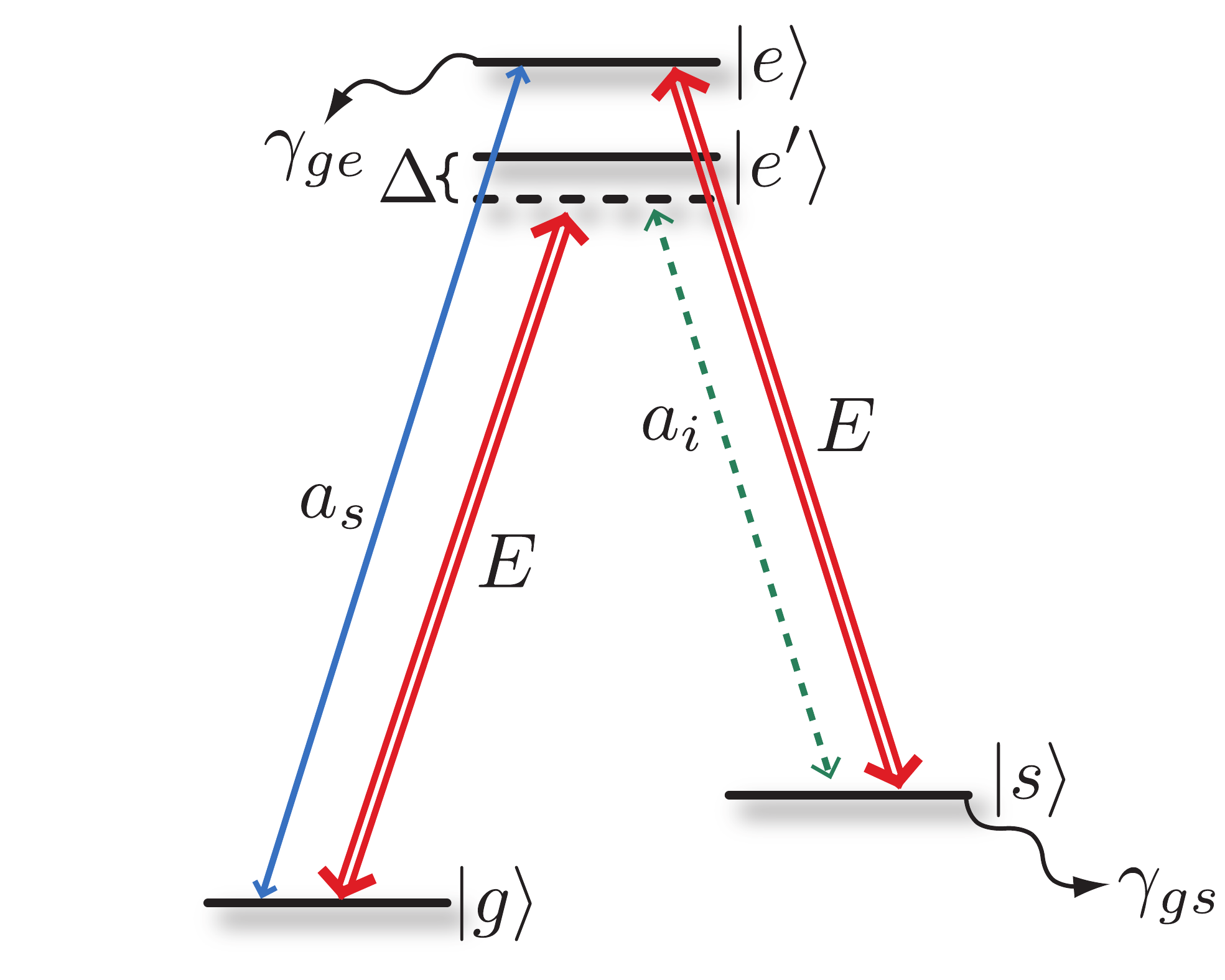}
\caption{The atomic level structure used for modeling EIT. A strong control field $E$ (red) is used to introduce coupling between the metastable $|s\rangle$ and the ground state $|g\rangle$. The $\Lambda$-system for EIT is formed by the control field and the signal field, $a_s$ (blue). A second $\Lambda$-system is formed by the control field and the idler field, $a_i$ (green), via another excited level $| e^{\prime}\rangle$, which completes a 4WM process. The Rabi frequency for both of the $\Lambda$-transitions are denoted by $\Omega$ and $\Omega^{\prime}$, respectively. The decay rates for the excited and the metastable states are given by $\gamma_{ge}$ and $\gamma_{gs}$, respectively.}
\label{fig:tla}
\end{figure}

To model the EIT-4WM process, we assume a four-level atom coupled by three optical fields as our model of the system as shown in fig.~\ref{fig:tla}. A strong control field $E$, with Rabi frequency $\Omega$, couples a meta-stable state $\ket{s}$, to the excited state $\ket{e}$ and a weak signal field (blue) $a_s$ couples the ground state $\ket{g}$ to $\ket{e}$, completing a two-photon transition between $\ket{g}$ and $\ket{s}$. The same control field also couples $\ket{g}$ to an auxiliary excited state $\ket{e'}$ off-resonantly, completing a second Raman transition with an idler field (green) $a_i$ that is generated by the four-wave mixing process. Adiabatically eliminating $\ket{e'}$, the equations of motion for this system are \cite{PhysRevA.88.013823}
\begin{align}
i \partial_t \sigma_{ge}  &= -i \gamma_{ge} \sigma_{ge} - g_s a_{s} - \Omega \sigma_{gs}  \label{Maxwell_Bloch:1} \\
i \partial_t \sigma_{gs}  &= -i \gamma_{gs} \sigma_{gs} - g_i (\Omega'/\Delta)  s^\dagger_{i} - \Omega^* \sigma_{ge} \label{Maxwell_Bloch:2} \\
(\partial_t + c \partial_z) a_{s} &= i g_s N \sigma_{ge} \label{Maxwell_Bloch:3} \\
(\partial_t + c \partial_z) a^\dagger_{i} &=  -i g_i N (\Omega'/\Delta) \sigma_{gs}, \label{Maxwell_Bloch:4} 
\end{align}
where $\sigma_{ge}$ and $\sigma_{gs}$ are collective spin-polarisation operators corresponding to the $\ket{g} \rightarrow \ket{e}$ and $\ket{g} \rightarrow \ket{s}$ transitions respectively in an ensemble of $N$ atoms. $\Delta$ is the frequency difference between $\ket{g}$ and $\ket{s}$ minus the frequency difference between $\ket{e}$ and $\ket{e'}$. The coupling rate for $a_{s(i)}$ is $g_{s(i)} $ and, similarly, $\Omega(\Omega^{\prime})$ is the Rabi frequency for the control field driving $\ket{s} \rightarrow \ket{e}$ ($\ket{g} \rightarrow \ket{e^{\prime}}$). $\gamma_{ge}$ and $\gamma_{gs}$ are the decay rates from $\sigma_{ge}$ and $\sigma_{gs}$. The rate $g_s$ differs from $g_i$ only due to the different dipole transition strengths associated with each transition, as does $\Omega$ from $\Omega^{\prime}$. We solve equations (\ref{Maxwell_Bloch:1}-\ref{Maxwell_Bloch:4}) in the Fourier domain to obtain the expression for the transfer function of the signal field in the absence of an injected idler. The output signal field, $a_s(z=L,t)$, after propagating through the ensemble of length L, is determined in terms of the input spectrum $a_s(z=0,\omega)$ by
\begin{widetext}
\begin{equation}
a_s(z=L,t) = \int T_s(\omega,z=L) a_s(z=0,\omega) e^{i\omega t} d\omega,
\label{outputpulse}
\end{equation}
\end{widetext}
with the transfer function $T_s(\omega)$ given by
\begin{widetext}
\begin{equation}
T_s(\omega) = e^{-\frac{D \gamma_{ge} }{ 4 V(\omega)} \left(i \omega -i \omega \vert\epsilon\vert^2+\vert\epsilon\vert^2\gamma_{ge}  -\gamma_{gs}\right)} \left(\frac{\left(\gamma_{ge} \vert\epsilon\vert^2-i \omega -i \vert\epsilon\vert^2\omega +\gamma_{gs}\right)}{ U(\omega)} \sinh\left[\frac{D\gamma_{ge} U(\omega)}{ 4 V(\omega)}\right]  +\cosh\left[\frac{D\gamma_{ge} U(\omega)}{ 4 V(\omega)}\right]\right),
\label{TransferF}
\end{equation}
\begin{align}
U(\omega) &= \sqrt{\left(i \omega +(i \omega -\gamma_{ge} ) \vert\epsilon\vert^2-\gamma_{gs}\right)^2+4 \vert\epsilon\Omega\vert^2} \nonumber\\
V(\omega) &= (i \gamma_{gs}+\omega ) (\omega +i \gamma_{ge} )-\vert\Omega\vert^2, \nonumber
\end{align}
\end{widetext}
with the definitions
\begin{align}
D = 2 \frac{g^2 N L}{\gamma_{ge}}; \qquad \epsilon &= \eta\frac{\Omega}{\Delta}; \qquad \eta = \frac{g_{i}\Omega'}{g_{s}\Omega} = \frac{d_{\ket{s} \rightarrow \ket{e'}} \cdot d_{\ket{g} \rightarrow \ket{e'}}}{d_{\ket{g} \rightarrow \ket{e}} \cdot d_{\ket{s} \rightarrow \ket{e}}}.
\end{align}
The parameter $\eta$ can be expressed in terms of the dipole matrix elements, $d_{\ket{j} \rightarrow \ket{k}}$, for the associated transition and the definition of optical depth, $D$, corresponds to an intensity attenuation of $e^{-D}$ when $\omega \rightarrow 0$ and $\Omega \rightarrow 0$.

\begin{figure}
\centering \includegraphics[width=0.6\columnwidth]{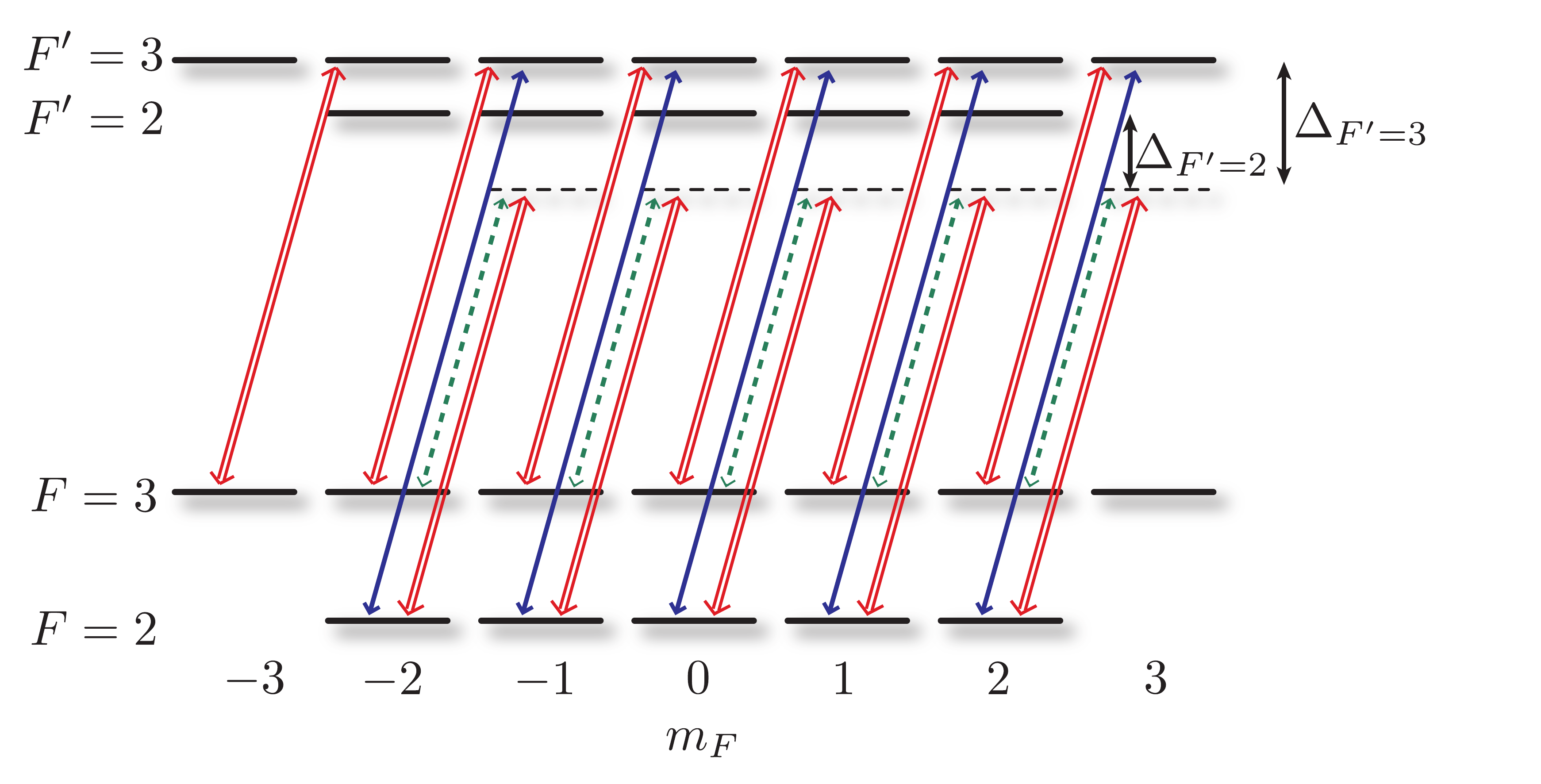}
\caption{The $^{85}$Rb D1 level structure used for the experiment shows five degenerate EIT-4WM systems. Each system differs by the Clebsch-Gordan coefficients associated with each transition.  The field labels are as shown in Fig.~\ref{fig:tla}}. 
\label{fig:Rb85Levels}
\end{figure}

The level structure of $^{85}$Rb atoms relevant for the EIT experiment is shown in Figure \ref{fig:Rb85Levels}, assuming $\sigma^+$ polarisations for all of the optical fields. We can identify five degenerate four-level structures coupling from each of the Zeeman sub-levels on the $F=2$ manifold. For each of the degenerate systems, both of the excited state manifolds ($\ket{F'=2}$ and $\ket{F'=3}$) are taken into account by summing the strengths of the off-resonant interaction with each excited state weighted by the relative detuning:
\begin{equation}
  \eta_{m_F} = \frac{d_{3,2} \cdot d_{2,2}}{d_{2,3} \cdot d_{3,3}} + \left( \frac{\Delta_{F'=2}}{\Delta_{F'=3}}\right) \frac{d_{3,3}\cdot d_{2,3}}{d_{2,3} \cdot d_{3,3}} 
  = \frac{d_{3,2} \cdot d_{2,2}}{d_{2,3} \cdot d_{3,3}} + \frac{\Delta_{F'=2}}{\Delta_{F'=3}},
\end{equation}
where $d_{i,j} \equiv d_{\ket{F=i, m_{F}} \rightarrow \ket{F'=j, m_{F+1}}}$ and $\Delta_{F'=j}$ is the detuning of the idler from the $F'=j$ excited state.

The degenerate EIT systems are reduced to a simple four-level model by the introduction of an effective interaction strength ration $\eta_\mathrm{eff}$. We assume that the population is uniformly distributed across the ground-state manifold, in which case $\eta_\mathrm{eff}$ can be approximated as the mean of the values of $\eta_{m_F}$ for each EIT system.  We find that for the D1 line of $^{85}Rb$ $\eta_\mathrm{eff}=1.62$.

\section{Experimental setup} \label{sec-setup}

\begin{figure*}
    \centering
    \includegraphics[width=\columnwidth]{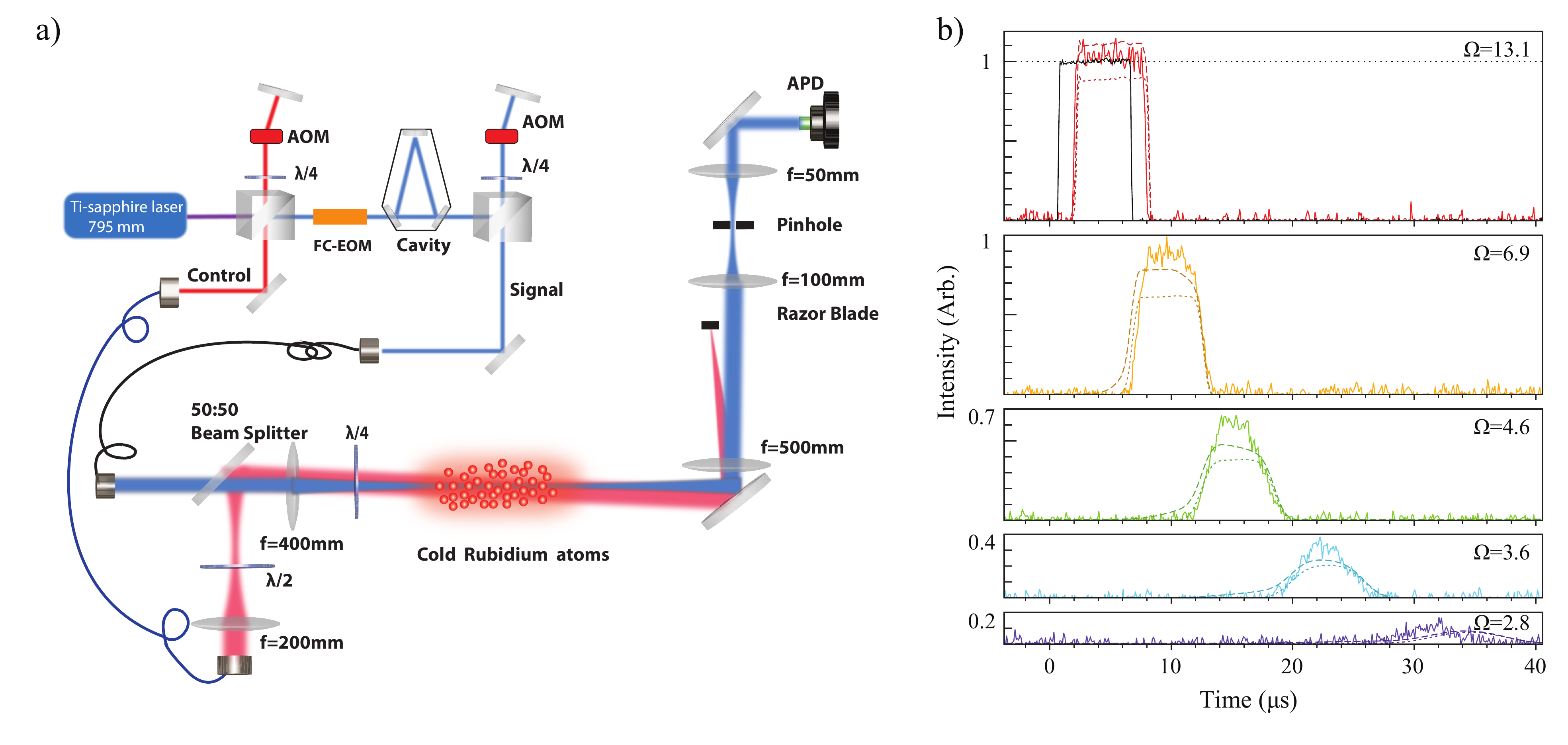}
    \caption{a) The experimental setup for EIT in a transiently compressed ensemble of cold ${}^{85}$Rb ($^{87}$Rb) atoms. The blue and red beams represented signal and control beams, respectively. The magneto-optical trap setup is similar to that of Ref.~\cite{Sparkes2013a} and is not explicitly shown here. b) Delayed pulses (solid, coloured traces) shown relative to the input (black) for 5 different signal field amplitudes.
Theoretical outputs, obtained by applying Eq.~(\ref{TransferF}) to the reference are shown as dashed lines. For comparison, theoretical outputs with no 4WM are shown as dotted lines.}
    \label{fig:setup}
\end{figure*}

The experiment was conducted using an elongated magneto-optical trap (MOT) for $^{85}$Rb ($^{87}$Rb) atoms \cite{Sparkes2013a}. Trapping was accomplished using two amplified external cavity diode lasers: one for cooling and one to optically pump the atoms back to the hyperfine state used for cooling. The cooling laser was 30 MHz red-detuned from the D$_2$ $F=3 \rightarrow F'=4$ ($F=2 \rightarrow F'=3$) transition and the repump is resonant with the  D$_2$ $F=2 \rightarrow F'=3$ ($F=1 \rightarrow F'=2$) transition. We loaded atoms for 480 ms, after which the MOT was compressed in two dimensions by smoothly increasing the transverse magnetic field gradients while simultaneously ramping down both the trapping frequency and intensity and the repump intensity over 20 ms \cite{Sparkes2013a}. Once the MOT was compressed the magnetic fields and the repump field were turned off. The trapping beams were turned off $50\mu$s after the repump field so that the atoms were pumped to the $F = 2$ ($F = 3$) ground state. We imposed a wait time of 500 $\mu$s to allow eddy currents to in the optical bench and other components to dissipate prior to turning on the signal field. Residual eddy currents continued to create a non-negligible magnetic field during the experimental window that varies at a rate of $\approx 3$ mG/$\mu$s.

The experimental setup is schematically shown in Figure~\ref{fig:setup}. The control field was produced by a Ti:Sapphire laser that was locked on resonance with the $^{85}$Rb ($^{87}$Rb) $D_1$ transition from $ F=3$ to $F'=3$ ($F=2 \rightarrow F'=2$). The signal field was produced by sending a portion of the Ti:Sapphire light through an electro-optical modulator (EOM) to produce modulation sidebands that were separated by $\sim3.035$ GHz ($\sim6.835$GHz) \cite{Steck85,Steck87}, the hyperfine splitting of $^{85}$Rb ($^{87}$Rb). The higher frequency of these sidebands was isolated using a filtering cavity and used as the signal field. Both the control and signal fields were gated using AOMs and both were $\sigma^+$ polarised. The signal pulse was focused along the long axis of the atom cloud with a beam waist of 200 $\mu$m to match the signal beam diameter to the cross section of the compressed atom cloud. The control beam was collimated to a diameter of 7 mm to ensure coverage of the entire atom cloud with uniform intensity and aligned to propagate with a small angle relative to the signal beam, overlapping it at the location of the MOT.

In order to measure the weak signal beam elimination of the control field was required. To accomplish this we employed two stages of spatial filtering. In the first stage, the control field was focused onto the edge of a razor blade which blocked the majority of the optical power. The second stage was a pinhole, through which the signal was focused, that served to eliminate most of the remaining scattered control field. The spatial filtering provided an extinction ratio of $\approx 45$ dB for the control while maintaining $\approx 80$\% signal detection efficiency.

\section{Results} \label{sec-results}

We examined the propagation of signal pulses through the atomic ensemble under the conditions of slow light. The signal pulse chosen as input had a square temporal profile with 6 $\mu$s width and was recorded after propagation through the ensemble for a variety of control field powers. A reference trace of the signal pulse shape was recorded by blocking the trapping beams so that the MOT was dispersed.

The predicted output pulse shape was calculated by applying the transfer function derived from the four-level model, Eq.~(\ref{TransferF}), to the recorded reference trace. A sample of traces recorded at an optical depth of $(550 \pm 20)$ is shown in Figure~\ref{fig:setup} (b) along with the theoretical output pulses. The values for optical depths used in the model were obtained using independent off-resonance absorption measurements. The only free parameters are the decay rate, $\gamma_{gs} = (12.8 \pm 0.5)$ kHz, and the ratio between $\vert\Omega\vert^2$ and the measured control field intensity which are fitted globally and consistent across all of the data. Other parameters can be found in Ref.~\cite{Steck85} with $\gamma_{ge}=\pi \cdot 5.75$MHz and $\Delta=2\pi \cdot 3.035$GHz. We found that there is some departure between the model and experiment in terms of the output pulse shape but that the total output power is well-predicted. We attribute this to pulse distortion caused by a residual time-dependent magnetic field resultant from eddy currents in the optical bench that persist after the trapping magnetic fields are turned off. This is based on the observation that adjustments to the background compensation field can partially correct the effect.

\begin{figure}
\centering \includegraphics[width=\columnwidth]{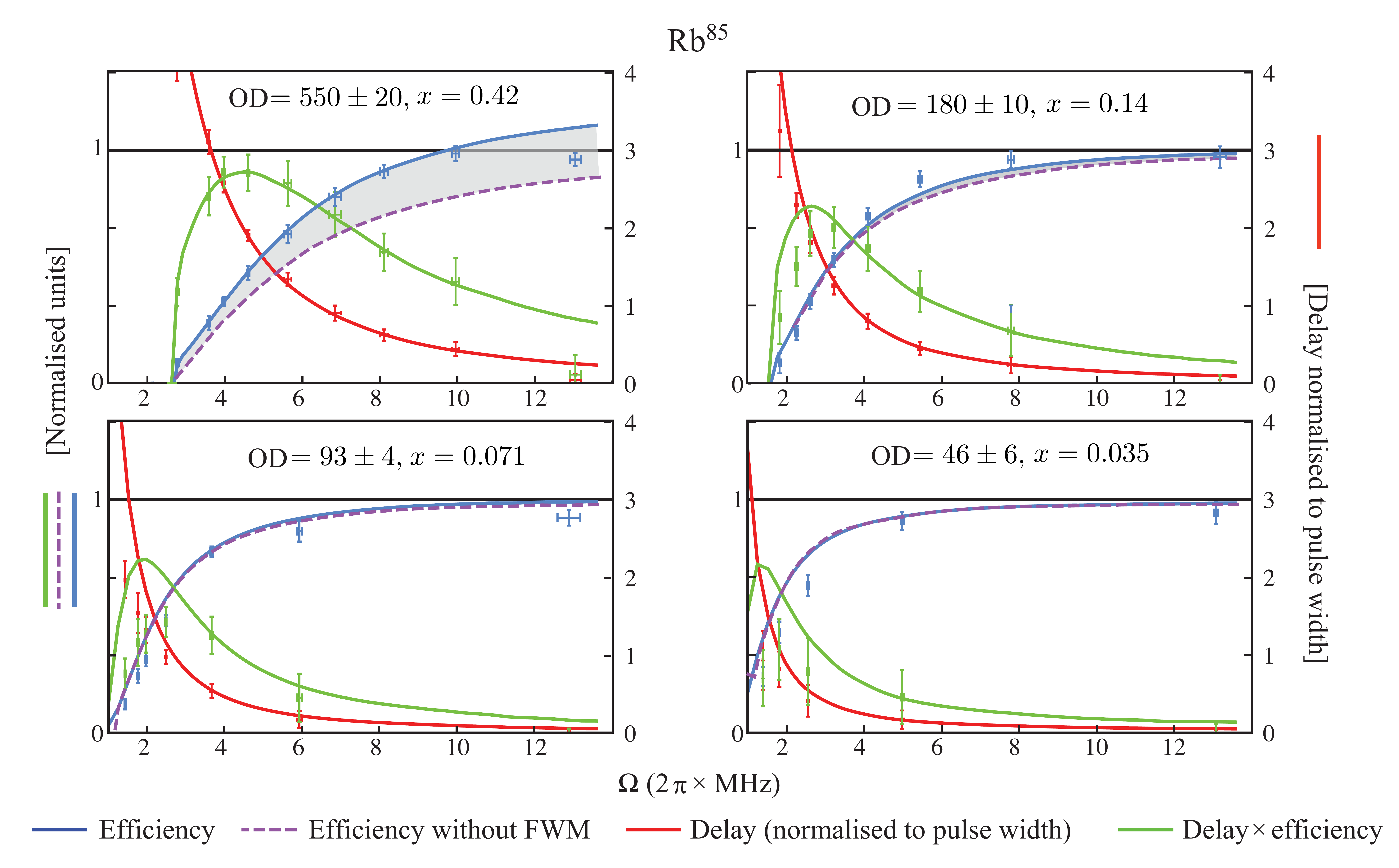}
\caption{The efficiencies (red), delay times  normalised to pulse width (purple) and the product of delay and efficiency (green) for the slow pulses relative to the reference input. The solid lines show the predicted values based on Eq.~(\ref{TransferF}). The dashed purple line shows the predicted efficiency if four-wave mixing is removed from the model.}
\label{fig:85efficiency}
\end{figure}
\begin{figure}

\centering \includegraphics[width=\columnwidth]{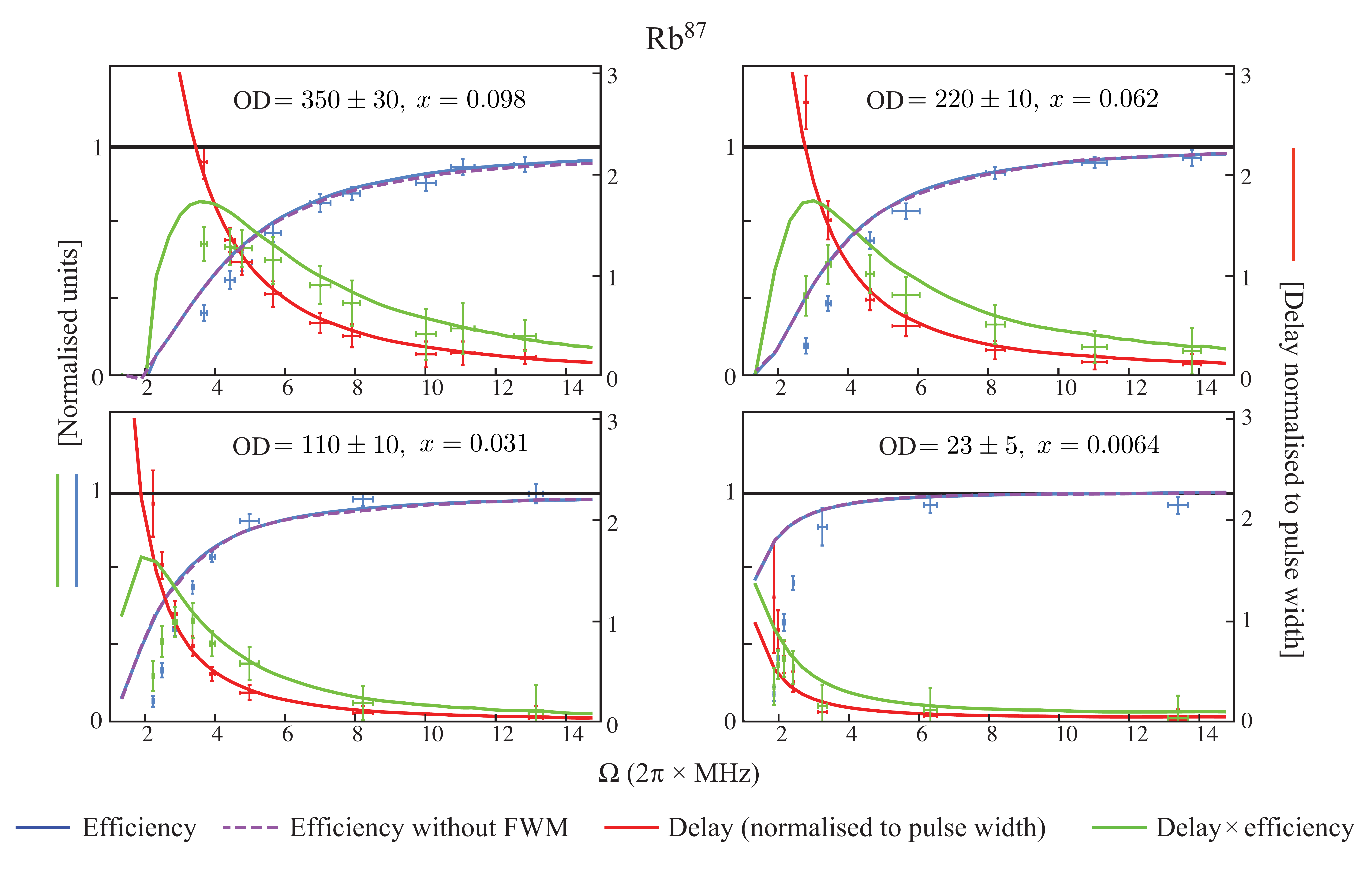}
\caption{The EIT efficiencies (blue), delay times  normalised to pulse width (red) and the product of delay and efficiency (green) for the slow pulses relative to the reference input. The solid lines show the predicted values based on Eq.~(\ref{TransferF}). The dashed purple line shows the predicted efficiency if four-wave mixing is removed from the model.}
\label{fig:87efficiency}
\end{figure}

The results demonstrate a gain in the signal after propagation through the slow-light medium, indicating that 4WM occured. In the noise analysis of Ref.~\cite{PhysRevA.88.013823}, the parameter
\begin{equation}
  x = \eta \frac{D}{2}\frac{\gamma_{ge}}{\Delta}
\end{equation}
is introduced to track the effective 4WM strength and we use it here to compare experiments in different Rb isotopes and at different optical depths. Values larger than 1 correspond to a regime where 4WM substantially impacts on the propagtion dynamics. While the effect is less pronounced for values less than 1, the contribution of noise photons can still be significant. At the highest optical depth achieved in our experiment, where OD $= 550 \pm 20$, the 4WM strength value was $x = 0.34$, corresponding to a regime where additional noise photons would have a detrimental effect on the fidelity of a quantum memory. Figure \ref{fig:85efficiency} shows the integrated output pulse intensities and delay times for a variety of control field powers at different optical depths. The total EIT efficiency and delay times are in good agreement with the model across the entire parameter space that we explored. For the two highest optical depths, theoretical EIT efficiency corresponding to zero 4WM strength are included. At lower optical depths, the theoretical predictions for both cases are indistinguishable with the resolution of the plot. The shaded region indicates the effect of 4WM on the efficiency of EIT. The observed gain is in good agreement with the simple four-level model, and contrasts with previous cold atom experiments that have speculated that four-wave mixing reduces the efficiency \cite{Zhang2011}. 

In contrast, an experiment conducted with $^{87}$Rb in the paper of Chen {\it et al.}~\cite{Chen2013} reported high storage efficiencies with negligible 4WM at an optical depth of 156. This discrepancy in observed 4WM is a result of the different effective 4WM strength in the experiment due to the lower optical depth and larger ground-state splitting in $^{87}$Rb relative to $^{85}$Rb. We verified that for $^{87}$Rb high delays are indeed possible with minimal 4WM by repeating the experiment with that isotope. While the maximum achieved optical depth of OD $= 350 \pm 30$, was lower than for $^{85}$Rb, high delay-bandwidth products were still obtained, as is shown in Figure~\ref{fig:87efficiency}. Here the effective interation strength ratio was calculated to have a theoretical value of $\eta_\mathrm{eff} = 1.33$, giving an effective 4WM strength of $x=0.08$ for an optical depth of 350. At the resolution of the plot, a theoretical line that includes 4WM is indistinguishable from one that does not have 4WM. Our results presented is consistent with the results of in Ref.~\cite{Chen2013}.

\begin{figure*}
\centering \includegraphics[width=\columnwidth]{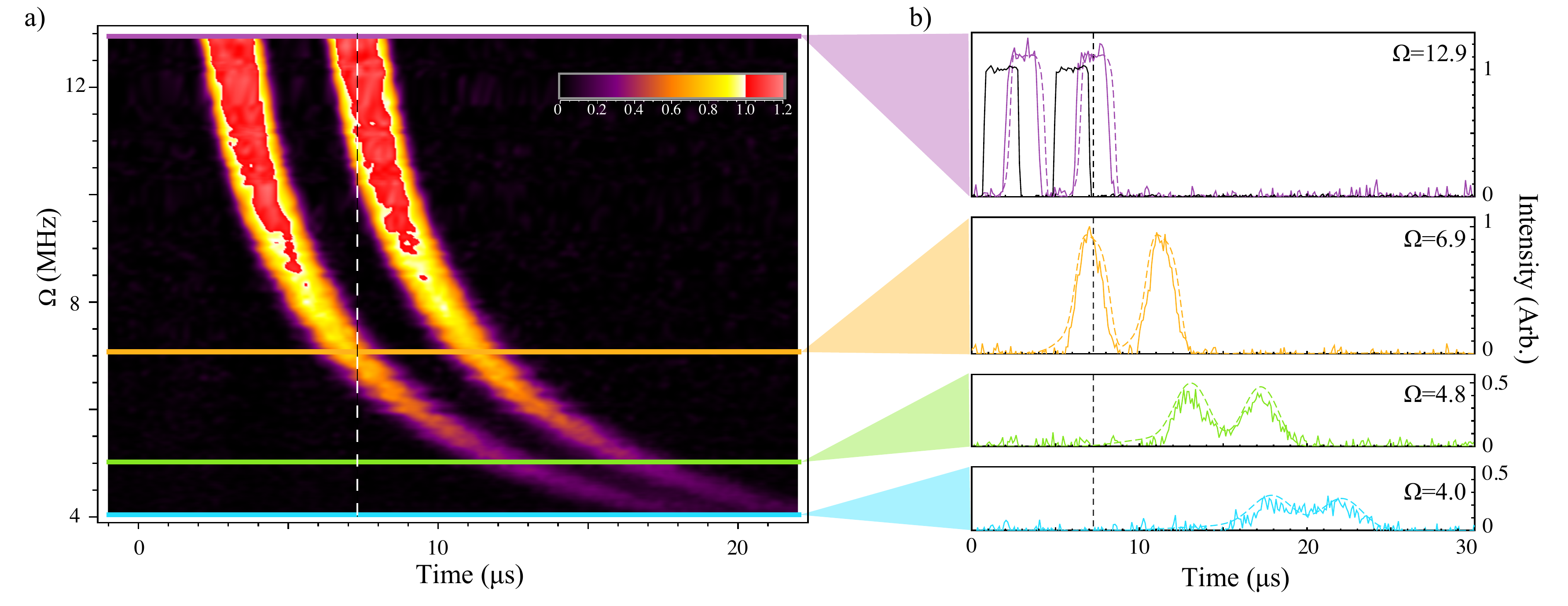}
\caption{Temporally multimode delay of two signal pulses. Here, the large optical depth is used to slow the input pulses enough that they are both contained entirely within the atomic ensemble. Part (a) shows the ouput from the delay medium as the control field power is varied. Individual traces from this data are shown in (b) with the associated traces highlighted in the same color in part (a). The top trace of (b) also shows the input pulse as a reference, with some 4WM gain being apparent in the output.}
\label{fig:two_pulse}
\end{figure*}

The multimodal capacity of EIT-based memory scales poorly with increasing optical depth; at best the modal capacity is $N \approx \sqrt{D}/3$ \cite{Nunn:2008}. In spite of this, the optical depth acheived in our experiment was sufficient to demonstrate enough delay that two pulses are contained entirely within the ensemble simultaneously. Figure \ref{fig:two_pulse} shows results taken at an optical depth of $(560 \pm 40)$ demonstrating a delay-bandwidth product of $\approx 3.7$, calculated by the ratio of the delay to transmitted pulse width at 50\% efficiency.

\section{Conclusion} \label{sec-concl}
We experimentally investigated slow light under the conditions of high optical-depth electromagnetically induced transparency in an ensemble of cold ${}^{85}Rb$ atoms. This is the first time that EIT has been observed at such large optical depths and we explored the role of four-wave-mixing in the EIT interaction. We found that a simple four-level model, one that has been used to theoretically predict the addition of 4WM noise  \cite{PhysRevA.88.013823}, was in good agreement with the experiment over a wide range of optical depths and in two Rb isotopes. This provides a solid foundation for predicting the noise performance of EIT-based optical qauntum memories.
	
In a regime with negligible 4WM, we obtained about 50\% efficiency with one pulse width delay. We additionally demonstrated a delay-bandwidth product of $\approx 3.7$ with 50\% efficiency. Although this corresponded to a regime with some 4WM, it enabled us to preform the first demonstration of multimode delay of temporally separated pulses in a cold-atom ensemble.
 
 \section*{ACKNOWLEDGMENTS}
We thanks Liqing Chen for the useful discussions. This work is funded by the Australian Research Council Centre of Excellence Program (CE110001027). WPZ acknowledge financial support from the National Basic Research Program of China (973 Program Grant No. 2011CB921604) and the National Natural Science Foundation of China (Grant Nos. 11234003). JG is supported by the Chinese Scholarship Council overseas scholarship. 

\bibliographystyle{unsrt}
\bibliography{EIT_Bib}

\end{document}